\documentclass{article}

\usepackage{amsfonts}
\usepackage{epsfig}

\begin{document}

\title{\textsc{Scaling and Universality in City Space Syntax: between Zipf and Matthew}}

\vspace{1cm}

\author{ D. Volchenkov and Ph. Blanchard
\vspace{0.5cm}\\
{\it  BiBoS, University Bielefeld, Postfach 100131,}\\
{\it D-33501, Bielefeld, Germany} \\
{\it Phone: +49 (0)521 / 106-2972 } \\
{\it Fax: +49 (0)521 / 106-6455 } \\
{\it E-Mail: VOLCHENK@Physik.Uni-Bielefeld.DE}}
\large

\date{\today}
\maketitle

\begin{abstract}

We report about  universality of rank-integration distributions 
of open spaces in city space syntax similar to 
the famous rank-size distributions of cities (Zipf's law). 
We also demonstrate that the
  degree of
choice an open space represents for
 other spaces directly linked to
it in a city  follows a power law statistic.
 Universal statistical behavior of space syntax measures
 uncovers the universality of
the city creation
mechanism. We suggest that the observed universality 
 may help 
to establish the international definition of a city
as a specific land use pattern.

\end{abstract}

\vspace{0.5cm}

\leftline{\textbf{ PACS codes: 89.75.Fb, 89.75.-k, 89.90.+n} }
 \vspace{0.5cm}

\leftline{\textbf{ Keywords: Complex networks, city space syntax} }

\section{Graphs and space syntax of urban environments}
\label{sec:Graphrepresentations}
\noindent

Urban space is of rather large scale
 to be seen from a single viewpoint;
maps provide us with its representations by
means of abstract symbols facilitating our
 perceiving and understanding of a city.
 The middle scale and small scale
maps are usually based on Euclidean geometry
 providing spatial objects with precise coordinates
 along their edges and outlines.

The widespread use of graph theoretic analysis in geographic
 science had been reviewed in \cite{Haggett}
establishing it as central to spatial analysis of urban environments.
In \cite{Kansky}, the basic graph
theory methods had been applied  to the measurements of
 transportation networks.

Network analysis has long been a basic function of geographic
information systems
(GIS) for a variety of applications, in which computational
 modelling of an urban network is based on a graph view in
 which the intersections of linear features are regarded as
nodes, and connections between pairs of nodes are represented
as edges \cite{MillerShaw}.
Similarly, urban forms are usually represented as the patterns
of identifiable urban elements such as
locations or areas (forming nodes in a graph) whose relationships
to one another are often associated with linear
transport routes such as streets within cities \cite{Batty}.
Such planar graph representations define locations or points in
Euclidean plane as nodes or vertices $\{ i\}$, $i=1,\ldots, N$, and
 the edges linking  them together as $i\sim j$, in
 which $\{i,j\}=1,2,\ldots,N.$ The value of a link can
 either be binary, with the value $1$ as $i\sim j$, and
 $0$ otherwise, or be equal to actual physical distance
 between nodes,  $\mathrm{dist}(i,j)$, or to some weight $w_{ij}>0$ quantifying
a certain characteristic property of the link.
We shall call a planar graph representing the Euclidean space
embedding of an urban network as its {\it primary graph}.
 Once a spatial system has been identified and
represented by a graph in this way, it can be subjected to
the graph theoretic analysis.

A {\it spatial network} of a city is a network of the spatial
 elements of urban environments. They
are derived from maps of {\it open spaces} (streets, places, and roundabouts).
Open spaces may be broken down into components; most simply, these
might be street segments, which can be linked into a network via
their intersections and analyzed as a networks of {\it movement
choices}. The study of spatial configuration is instrumental in
predicting {\it human behavior}, for instance, pedestrian
movements in urban environments \cite{Hillier96}. A set of
theories and techniques for the analysis of spatial configurations
is called {\it space syntax} \cite{Jiang98}. Space syntax is
established on a quite sophisticated speculation that the
evolution of built form can be explained in analogy to the way
biological forms unravel \cite{SSyntax}. It has been developed as
a method for analyzing space in an urban environment capturing its
quality
 as being comprehendible and easily navigable \cite{Hillier96}.
Although,  in its initial form, space syntax was focused mainly on
patterns of pedestrian movement in cities, later the  various
space syntax measures of urban configuration had been found to be
correlated with the different aspects of social life,
\cite{Ratti2004}.

Decomposition of a space map into a complete set of
intersecting axial lines,  the fewest and
longest lines of sight that pass through every open space comprising any system,
produces an axial map or an overlapping convex map respectively.
Axial lines and convex spaces may be treated as the {\it spatial elements}
 (nodes of a morphological graph),
 while either the {\it junctions} of axial lines or the {\it overlaps} of
 convex  spaces may be considered as the edges linking  spatial elements
 into a single  graph unveiling the
topological relationships  between all open elements of the urban space.
In what follows,  we shall call this morphological representation of urban network
as a {\it dual graph}.

The encoding of cities into non-planar dual graphs
reveals their complex
structure.
The transition to a dual graph is a topologically non-trivial
transformation of a planar primary graph into a non-planar one which
encapsulates the hierarchy and structure of the urban area and also corresponds
 to perception of space that people experience when
travelling along routes through the environment.
The dual transformation replaces the 1D open segments (streets) by the
 zero-dimensional nodes.
The sprawl like developments consisting of a number of blind passes branching
 off a main route are changed to the star subgraphs having a hub and a number
 of client nodes. Junctions and crossroads are replaced with
 edges connecting the corresponding nodes of the dual graph.
  Places and roundabouts are considered as the independent topological
  objects and acquire the individual IDs being nodes in the dual
  graph. Cycles are converted into  cycles of
  the same lengthes.
The dual graph representation of a regular grid pattern
is a {complete bipartite} graph, where the set of vertices
  can be divided into two disjoint subsets  such that no edge has
   both end-points in the same subset, and every line joining the
   two subsets is present, \cite{Krueger89}.
These sets can be naturally interpreted as those of the vertical
and horizontal edges in the primary graphs (streets and avenues).
 It is the dual
graph transformation which allows to separate the effects of order
and of structure while analyzing a transport network on the
morphological ground. It converts the repeating geometrical
elements expressing the order in the urban developments into the
{\it twins nodes}, the pairs of nodes such that any other is
adjacent either to them both or to neither of them.

{\it Integration} is a centrality measure used in space syntax theory 
in order to 
 express the
degree to which a node is integrated or segregated from
the whole urban texture, \cite{JAG}.
In the present paper, we report on the Rank-Integration distributions
observed 
for the several compact urban patterns
similar to the famous Rank-Size distribution ("Zipf's law") 
(see Sec.~\ref{sec:ControlChoice_statistics}). 
Furthermore, we show that 
the {\it control value} parameter used in 
space syntax theory for the evaluation of the
 {\it degree of
choice} each node represents for
 nodes directly linked to
it follows a power law statistic in the intermediate range of scales (see Sec.~\ref{sec:ControlChoice_statistics}).
Universal statistical behavior of space syntax measures
 uncovers the universality of
the creation
mechanism responsible for the
 appearance of
nodes of high centrality which acts over
all cities independently
of their backgrounds. 
In Sec.~\ref{subsec:CAMEO},\ref{subsec:Sprawls}, we
discuss the possible mechanicsms of 
city formation which may be
responsible for the appearance of universality in all 
studied compact urban patterns. 
 
We have studied
 five different compact urban patterns.
 Two of them are
situated on islands: Manhattan (with an almost regular grid-like
city plan) and the network of Venice canals
(imprinting the joined
effect of natural, political, and economical factors acting on the
network during many centuries).
In the old city center
of Venice that stretches across 122 small islands in the marshy
 Venetian Lagoon along the Adriatic Sea in northeast Italy,
 the canals serve the function of roads.

We have also considered two organic cities
founded shortly after the Crusades and developed within the medieval
fortresses: Rothenburg ob der Tauber, the medieval Bavarian city
preserving its original structure from the 13$^\mathrm{th}$ century,
and the downtown of
Bielefeld  (Altstadt Bielefeld),
 an economic and cultural center of Eastern Westphalia.

To supplement the study of urban canal networks, we have
investigated that one in the city of Amsterdam. Although
it is not actually isolated from the national canal network, it is
binding to the delta of the Amstel river, forming a dense canal
web exhibiting a high degree of radial symmetry.

The scarcity of
physical space is among the most important factors determining the
structure of compact urban patterns. Some characteristics of studied dual city graphs are
given in Tab.~1. There, $N$ is the number of open spaces (streets/canals and places)
 in the urban pattern
(the number of nodes in the dual graphs), $M$ is the number of junctions (the number of
edges in the dual graphs);
the graph {\it diameter},
$\mathrm{diam}(\mathfrak{G})$ is the {\it maximal} depth (i.e., the graph-theoretical distance)
between two vertices in a dual graph.

\section{Space syntax measures}
\label{sec:Analyticmeasures}
\noindent

In space syntax theory,  graph-based models
 of space are used in order
to investigate the influence of the
shape
and configuration of environments on
human spatial behavior and experience. 
A number of configurational {\it measures}
have been introduced in so far
in quantitative representations
of relationships between
spaces of urban areas and buildings.
Below we give a
brief introduction into
the measures commonly accepted in
 space syntax theory.

Although similar parameters
quantifying
connectivity and centrality  of nodes in a graph
have been independently invented and
extensively studied during
the last century in a varied range of disciplines
including
computer science, biology, economics, and sociology, the
syntactic measures are by no means just the new names for the well
known quantities.
In space syntax, the spaces are
understood as voids between
buildings restraining
traffic  that dramatically changes their meanings and the
interpretation of results.

The main focus of the space syntax study is on the
relative proximity (or {\it accessibility}) between different locations
which involves calculating graph-theoretical {\it distances} between
nodes of the dual graphs and associating these distances with densities and
intensities of human activity which occur at different open spaces and
along the links which connect them \cite{Hansen59,Wilson70,Batty}.

Space {\it adjacency} is a basic rule to form axial maps:
two axial lines
 intersected are regarded as adjacency. Two spaces, $i$ and $j$, are
held to be {\it adjacent} in the dual graph $\mathfrak{G}$
when it is possible
to {\it move freely} from one
space to another, without passing through any
intervening.

The {\it adjacency matrix}  ${\bf A}_\mathfrak{G}$ of the dual graph
   $\mathfrak{G}$ is defined as follows:
\begin{equation}
\label{adjacencymatrix}
({\bf A}_\mathfrak{G})_{ij}=\left\{
\begin{array}{ll}
1,& i\sim j, \\
0,& \mathrm{otherwise}.
\end{array}
\right.
\end{equation}
Let us note that
rows and columns of ${\bf A}_\mathfrak{G}$
corresponding to
 the twins nodes are identical.
{\it Depth} is a topological distance
between nodes
in the dual graph $\mathfrak{G}$.
Two open spaces,  $i$  and $j$, are said to
 be at  depth $d_{ij}$ if
the
{\it least number} of syntactic steps
needed to reach one node from the other
is $d_{ij}$, \cite{glossary}.
The concept of depth can be extended to {\it total depth},
the sum of all
depths from a given origin,
\begin{equation}
\label{total_depth}
\mathfrak{D}_i=\sum_{j=1}^N d_{ij},
\end{equation}
in which $N$ is the total number of nodes in  $\mathfrak{G}$.
The average number of syntactic
steps from a given node $i$ to any other node
in the dual graph $\mathfrak{G}$
is called the
{\it mean depth},
\begin{equation}
\label{meandepth}
\ell_i=\frac {\mathfrak{D}_i}{N-1}.
\end{equation}
 The mean depth (\ref{meandepth}) is
 used for
quantifying the level of integration/segregation of the given
node,
\cite{Jiang98}.

{\it Connectivity}
is defined in space syntax theory as the number of nodes that
connect directly to a given node in the dual graph $\mathfrak{G}$, \cite{glossary}.
In graph theory, the space syntax connectivity\footnote{In
 graph theory, connectivity of a node is defined as the
{\it number of edges} connected to a vertex. Note that it
  is not
necessarily equal to the degree of node,  $\deg(i)$,  since there may be
more than one edge between any two vertices in the graph.}
 of a node is called the
 node  {\it degree}:
\begin{equation}
\label{connectivity}
\begin{array}{lcl}
\mathrm{Connectivity}(i) & =& \deg(i)\\
                         & = & \sum_{j=1}^N({\bf A}_\mathfrak{G})_{ij} .
\end{array}
\end{equation}
The {\it accessibility} of a space is considered in space syntax
as a key determinant of its spatial interaction and its analysis
is based on an implicit graph-theoretic view of the dual graph.

{\it Integration} of a node
is by definition expressed by a value that indicates the
degree to which a node is integrated or segregated from
a system as a whole ({\it global integration}), or from a partial
system consisting of nodes a few steps away ({\it local
integration}), \cite{JAG}.
It is measured by the {\it Real Relative Asymmetry} (RRA) \cite{Krueger89},
\begin{equation}
\label{integration2}
\mathrm{RRA}(i)\,=\, 2\,\frac{\ell_i -1}{D_N\,(N-2)},
\end{equation}
in which the normalization parameter
allowing to compare nodes belonging
to the dual graphs of different sizes
 is
\begin{equation}
\label{D_value}
D_N= 2\frac{N\left(\log_2\left(\frac{N+2}{3}\right)-1\right)+1}
{(N-1)(N-2)}.
\end{equation}
Another local measure used in  space syntax theory is the
{\it control value} (CV). It evaluates the degree
to which a space controls access to its immediate
neighbors taking
into account the number of alternative connections that each of
these neighbors has. The control value is
determined according to the following
formula, \cite{Jiang98}:
\begin{equation}
\label{controlvalue}
\begin{array}{lcl}
\mathrm{CV}(i) &=& \sum_{i\sim j}\frac 1{\deg(j)}\\
 &=& \sum_{j=1}^N\left({\bf A}_\mathfrak{G}{\bf D}^{-1}\right)_{ij}\\
\end{array}
\end{equation}
where the
diagonal matrix is
$
{\bf D}=\mathrm{diag}\left(\deg(1),\deg(2),\ldots,\deg(N)\right).
$

A dynamic global measure
of the {\it flow} through a space $i\in \mathfrak{G}$ commonly accepted
in  space syntax theory  is the
{\it global choice}, \cite{Hillier1987}. It captures how often a node may
be used in journeys from all spaces to all others
spaces in the city.
Vertices that occur on many shortest paths between other vertices have higher
 betweenness than those that do not.
Global choice can be estimated as the ratio between the number
 of shortest paths through the node $i$ and the total number of
  all shortest paths in $\mathfrak{G}$,
\begin{equation}
\label{globalchoice}
\mathrm{Choice}(i)=
\frac{\{\# \mathrm{ shortest {\ } paths {\ } through {\ }} i \}}
{\{\# \mathrm{ all {\ }shortest {\ } paths\}}}.
\end{equation}
A space $i$ has a
{\it strong choice} value when many of the shortest
paths, connecting all spaces to all spaces of a
system, passes through it.
 The Dijkstra's
classical  algorithm
which visits all nodes that are closer to the source
than the target before reaching the target
can be implemented in order to compute the value $\mathrm{Choice}(i)$.

The integration and the global choice index are the
centrality measures which capture the relative structural importance of a
node in a dual  graph.

\section{Scaling and universality in city space syntax}
\label{sec:ControlChoice_statistics}
\noindent

In urban studies, scaling and universality have been found  with
a remarkable regularity.

The famous {\it rank-size distribution}
of city sizes all over the world is known as {\it Zipf's Law}.
If we calculate the natural logarithm of the city rank in some
countries\footnote{The
empirical validity of Zipf's Law for cities
have been checked out recently
using new data on the city populations over  73
countries by two different estimation methods in \cite{ZipfSoo}.
 The use of various estimators justifies the validity of Zipf's
 Law for from 20 to 43 of 73 investigated countries. It has been
 suggested in \cite{ZipfSoo} that  variations in the value of
 Zipf's exponent are better explained by
political economy variables than by economic geography
 variables.} and of the city size (measured in terms of its population)
and then plot the resulting data in a diagram,  we obtain a remarkable
linear pattern with the slope of the line
 equals $-1$ (or $+1$, if cities have been ranked in the ascending  order),
\cite{ZipfSoo}.

  A possible explanation for the rank-size
distributions of the human settlements based on simple stochastic
models of settlement formation and growth has been recently proposed in
\cite{Reed}.

\begin{figure}[ht]
\label{Fig1_16a}
 \noindent
\begin{center}
\epsfig{file=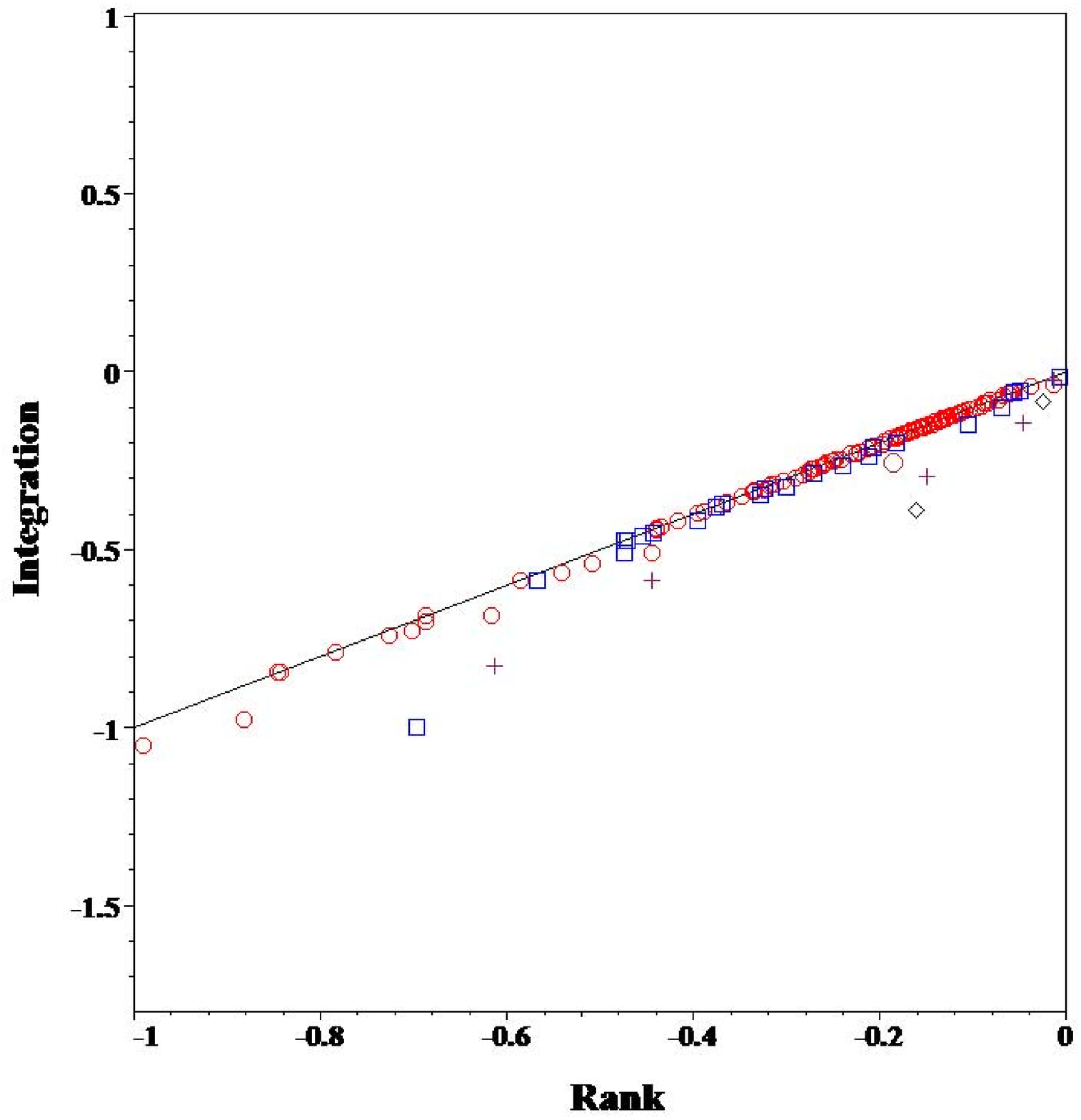, width=7.0cm, height =7cm}
\end{center}
\caption{The log-log plot of the Rank-Integration distribution
calculated the nodes of five
 dual graphs: the street grid of Manhattan, Rothenburg,
  Bielelefeld, the canal networks in Venice and Amsterdam.
    The linear pattern
    is given by a straight line. }
\end{figure}

Here we report on the similar rank distributions for the values of
the space syntax measures quantifying the {\it centrality} of
nodes in the dual graphs calculated for the compact urban
patterns.

We have ranked all open spaces in the dual graphs of the investigated
 five compact urban patterns in accordance to their centrality indices.
The nodes with the worse centrality have been assigned to the first (lowest)
rank, while to those of the best centrality the highest rank has been given.
 The diagrams for {\it Rank-Centrality } distributions have been calculated for
  all dual graphs representing
two German medieval cities (Rothenburg of der Tauber and
Bielefeld), the street grid in Manhattan, and two city canal
networks, in Amsterdam and in Venice, within the same frame. It is
worth to mention that no matter how the centrality level was
estimated, either by the integration values (\ref{integration2})
 or by the global choice
values (\ref{globalchoice}), the data from all
centrality
 indicators demonstrate a surprising {\it universality}
being fitted perfectly with the linear pattern, with the slope of the line equals $1$.

\begin{figure}[ht]
\label{Fig1_16b}
 \noindent
\begin{center}
 \epsfig{file=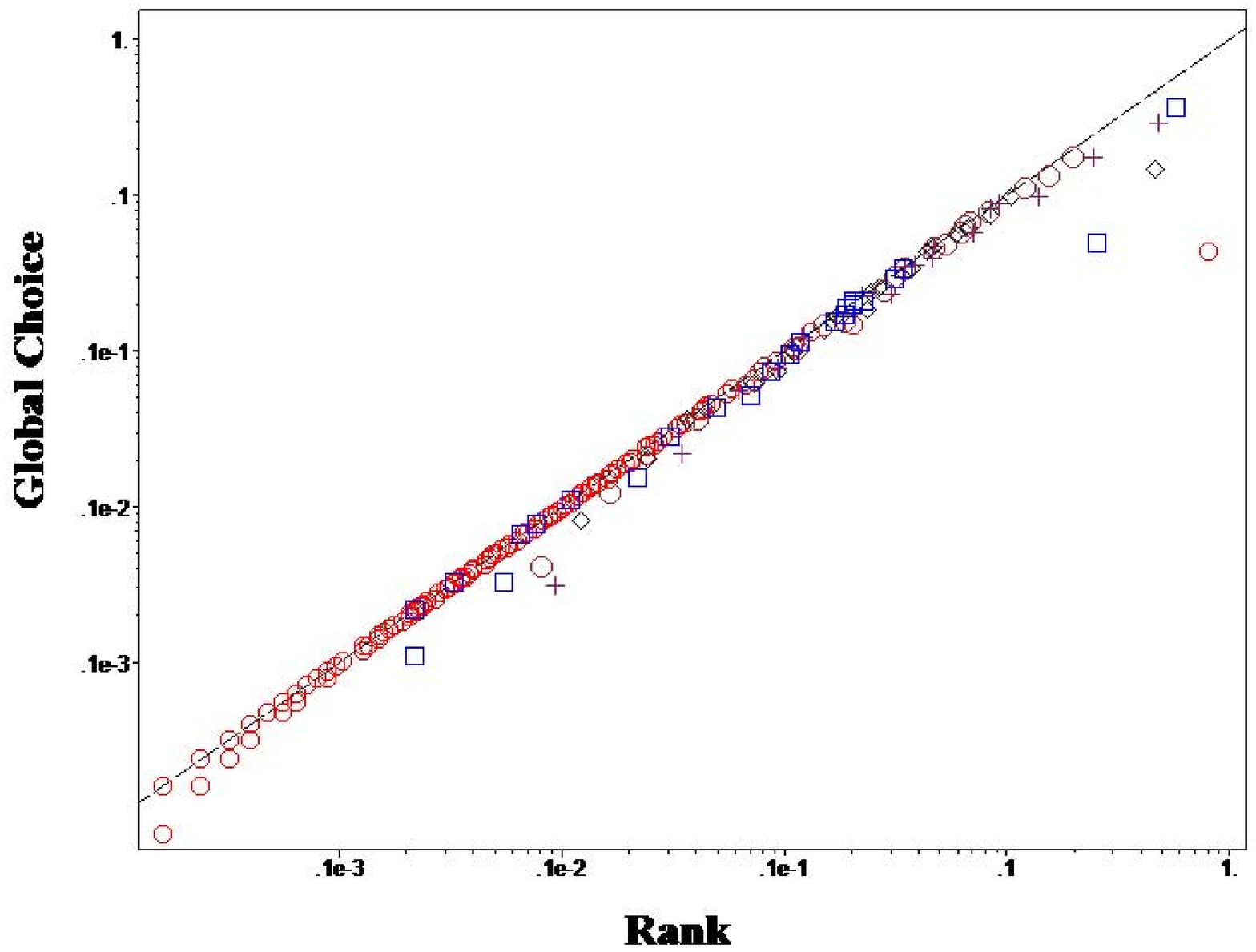, width=7.0cm, height =7cm}
\end{center}
\caption{ The log-log plot of the Rank-Global
    Choice (Betweenness) distribution calculated the nodes of five
 dual graphs: the street grid of Manhattan, Rothenburg,
  Bielefeld, the canal networks in Venice and Amsterdam.
    The linear pattern
    is given by a straight line. }
\end{figure}

The matching of power law behaviors
 (the same scaling
exponent)
can have a deeper origin in the
background dynamical process
responsible for such a
power-law relation. Being diverse in their sizes, forms,
economical and political history, cities  display nevertheless the
identical {\it scaling behavior} and probably share the similar
fundamental mechanisms of the open spaces creation. Formally, such
a common dynamics can be referred to as {\it universality}, so
that those cities with precisely the same critical exponents of
rank-integration statistics are said to belong to the same {\it
universality class}.

The ubiquity
of power-law relations in
complex systems are often thought
to be signatures of hierarchy
and robustness.
The area
distribution of satellite cities around large urban centers has
been reported to obey a power-law with exponent $\simeq 2$,
\cite{Stanley}.
The fractal dimension of urban aggregates as a global measure
of areal coverage have been studied extensively for many cities
around the world during the last decades
(see \cite{FractalCities},\cite{Tannier}
 for a review).
 The scaling property has also been observed
  recently
in concern with the space syntax studies, in the distribution
 of the length of open space linear segments (axial lines)
  \cite{Carvalho}.

In the present subsection, we discuss a scaling property
of control
   values distribution
calculated for the dual graphs of compact urban patterns.

The $\mathrm{CV}(i)$-parameter
quantifies the degree of choice the node $i\in\mathfrak{G}$
represents for other
nodes directly connected to it.

\begin{figure}[ht]
 \noindent
\begin{center}
\epsfig{file=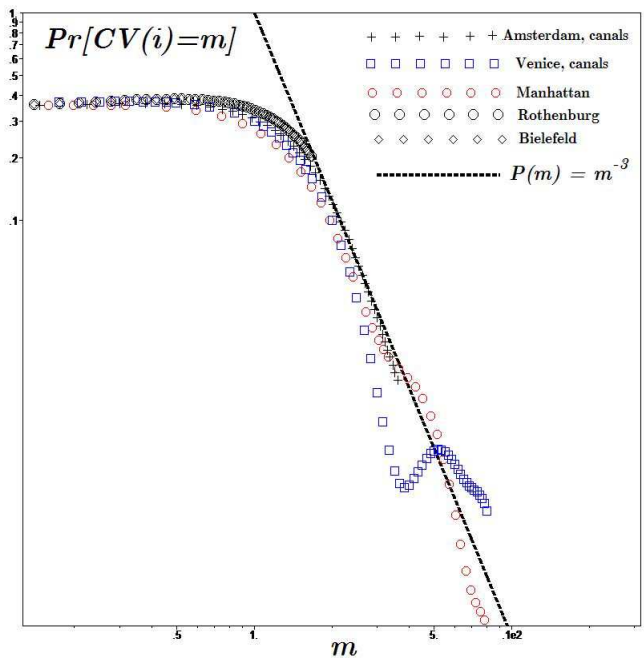, angle=0,width =9cm, height =9cm}
  \end{center}
\caption{\small The log-log plot of the probability distribution
that a node randomly selected among all nodes of the dual graph
 $\mathfrak{G}$ will be populated with precisely $m$ random walkers
  in one step starting from the uniform distribution (one random
   walker at each node). The dashed line indicates the cubic hyperbola
    decay, $P(m)=m^{-3}$.}
\label{Fig1_15}
\end{figure}

Provided random walks,
in which a  walker
moves in one step to
another node randomly chosen
 among all its nearest neighbors
 are
defined on the graph
 $\mathfrak{G}$, the
parameter $\mathrm{CV}(i)$
acquires a probabilistic interpretation.
Namely, it specifies
 the expected
number of
  walkers which is found in $i\in\mathfrak{G}$
after one step
if the random walks starts
 from a uniform configuration, in which
all nodes in the graph
have been  uniformly
populated
by precisely one walker.

Then, a graph $\mathfrak{G}$
can be characterized
by the probability
\begin{equation}
\label{CVdistr01}
P(m)=\Pr\left[i\in \mathfrak{G}|\mathrm{CV}(i)=m\right]
\end{equation}
of that the control value of a node chosen
uniformly at random among all nodes of the graph
$\mathfrak{G}$ equals to $m>0$.

The log-log plot of (\ref{CVdistr01}) is shown in  Fig.~\ref{Fig1_15}.
It is important to mention that
the profile of the probability decay
 exhibits the approximate scaling
well fitted by the cubic hyperbola,
  $P(m)\simeq m^{-3}$,
universally for all five compact urban
 patterns we have studied.

Universal statistical behavior of the
 control values for the nodes
representing a relatively
strong choice
for their nearest neighbors,
\begin{equation}
\label{CVstat}
\Pr\left[i\in \mathfrak{G}|\mathrm{CV}(i)=m\right] \simeq \frac 1{m^3},
\end{equation}
 uncovers the universality of
the creation
mechanism responsible for the
 appearance of
the "strong choice"
nodes which acts over
all cities independently
of their backgrounds.
 It
is a common suggestion in  space syntax theory
that  open spaces of strong choice
are
responsible for the {\it public space processes}
driven largely by the
universal social activities like trades
and exchange
which are common
across different cultures
and historical epochs
and  give cities
a similar global
structure of the "deformed wheel" \cite{SSBeijing}.

It has been shown long time ago by H. Simon \cite{Simon}
that the
power law distributions always arise when the {\it "the rich get richer"}
principle works, i.e. when a quantity increases with its amount already present.
In sociology this principle is known as the {\it Matthew effect} \cite{Matthew}
(this reference appears in \cite{Newman2003}) following the well-known biblical
edict.
In the next section, we discuss the graph evolution algorithms related to
"the  rich get richer" principle.

\subsection{Cameo principle of scale-free urban  developments}
\label{subsec:CAMEO}
\noindent

It is interesting to consider
possible city development algorithms that could
lead to the emergence of scaling invariant
degree structure in urban environments.

Among the classical models in which the degree distribution
of the arising graph
satisfies a power-law is
the graph generating algorithms
based on  the {\it preferential attachment} approach.
Within preferential
   attachment algorithms, the growth of a network starts
   with an initial graph of $n_0\geq 2$ nodes such that
   the degree of each node in the initial network is at
   least 1.
The celebrated Barab\'{a}si-Albert model \cite{BA} have been
 proposed in order to model the emergency and growth
  of scale-free complex networks.
 New nodes in the model \cite{BA} are
added to the network one at a time. Each
 new node is connected to $n$ of the existing with a
  probability that is biased being proportional
   to the number of links that the existing node already has,
\begin{equation}
\label{Bas}
p_i=\frac {\deg(i)}{\sum_{j=1}^N\deg(j)}.
\end{equation}
It is clear that the
nodes of high degrees tend to quickly accumulate even more
 links representing a strong {\it preference choice} for
 the emerging nodes, while nodes with only a few links
 are unlikely to be chosen as the destination for a new
 link.
The preferential attachment forms
 a positive feedback
loop in which an initial random degree variation is magnified with
time, \cite{BA2}. It is fascinating that the expected degree
distribution in the graph generated in accordance to the algorithm
proposed in \cite{BA} asymptotically approaches the cubic
hyperbola,
\begin{equation}
\label{cubichyperbola}
\Pr\left[i\in \mathfrak{G}|\deg(i)=k\right]\simeq \frac 1{k^3}.
\end{equation}
It is however obvious that
the mechanisms
governing the city creation and development
certainly do not
follow such a simple preferential
 attachment principle as that discussed in \cite{BA}.
Indeed, when new streets (public or private) are created as a
result of site subdivision or site redevelopment, they can hardly
be planed in such a way as to meet
 the streets
that already have the ever
maximal number of junctions
with other streets in a city.
The challenge of
city modelling calls for
the more realistic heuristic principles
that could catch the main features
of city creation and development.

It is clear that a prominent model
describing the urban developments should
take into account
the structure of embedding physical space:
the size and shape of landscape, and the
 local land use patterns.
A suitable algorithm
describing the
 development of
complex networks which
takes into account
has been recently  proposed
in \cite{CAMEO}.
It is called the {\it Cameo-principle} having
in mind the attractiveness,
rareness and beauty of the small medallion with a
 profiled head in relief called
Cameo. It is exactly their rareness and beauty which
gives them their high
value.

In the  Cameo model \cite{CAMEO}, the local attractiveness of a
site determining the creation of new spaces of motion in that is
specified by a real positive parameter $\omega >0$. Indeed, it is
rather difficult if ever be possible to estimate exactly
 the actual value $\omega(i)$ for any
 site $i\in \mathfrak{G}$
in the urban pattern, since such an estimation can be referred to
both the {\it local believes} of city inhabitants
 and may be to the {\it cultural context}
 of the site
 that may
vary over the
different nations, historical epochs, and
even over the certain groups of population.

Therefore, in the framework of the probabilistic approach,
 it seems natural to consider the
value $\omega$ as a real positive
 independent {\it random variable}
distributed over the
vertex set of the graph representation of the
site uniformly
 in accordance to
a smooth monotone decreasing probability
density function
$f \left( \omega \right)$.

Let us suggest
that there is just a few
distinguished
sites which are much more
attractive then an
average one in the city,
so that the density function $f$
has a right tail for
large $\omega\gg \bar{\omega}$ such that
  $f(\omega)\ll f(\bar{\omega})$.

Each newly created space of motion $i$ (represented by a node in
the dual city graph $\mathfrak{G}(N)$ containing $N$ nodes) may be
arranged in such a way to connect to the already
 existed space
$j\in\mathfrak{G}(N)$
depending only on its attractiveness
 $\omega(j)$ and is of the form
\begin{equation}
\Pr\left[\,
i\sim j \,\,|  \,\, \omega(j) \, \right]
\simeq
\frac{1}{N\cdot f^{\alpha}\left(\omega(j)\right)}
\label{P1.C1.01}
\end{equation}
\noindent
with some $ \alpha \in \left( 0,1\right)$.  The assumption
(\ref{P1.C1.01}) implies that the probability to create
the new space adjacent to a space $j$ scales with
the rarity of sites characterized with the same
attractivity $\omega$ as $j$.

The striking observation under the above assumptions is the
emergence of a scale-free degree distribution independent of the
choice of distribution $f(\omega)$. Furthermore, the exponent in
the asymptotic degree distribution
becomes independent of the distribution $f(\omega)$ provided its tail,
  $f(\omega)\ll f(\bar{\omega}),$
 decays faster then any power law.

In the model of growing networks proposed  in \cite{CAMEO}, the
initial graph $\mathfrak{G}_0$ has $N_ 0$ vertices, and a  new
vertex of attractiveness $\omega$ taken independently uniformly
 distributed in accordance to
 the given
density $f(\omega)$ is added to the already existed network at
each time click $t\in \mathbb{Z}_+$. Being associated to the
graph, the vertex establishes $k_0>0$ connections with other
vertices already present in that. All edges are formed accordingly
to the Cameo principle (\ref{P1.C1.01}).

The main result of \cite{CAMEO}
is on the
probability distribution
that a randomly chosen vertex $i$
which
had joined the Cameo graph $\mathfrak{G}$
at time $\tau>0$
 with  attractiveness $\omega(i)$
amasses
precisely $k$ links from other vertices
which emerge by time $t>\tau$.
It is
important to note that
in the Cameo model the order in which the
edges are created plays a
role for the fine structure of the graphs.
The resulting degree distribution
for $t-\tau> k/k_0$
is irrelevant to
the
concrete form of $f(\omega)$ and reads as following
\begin{equation}
\label{Krueger}
\Pr\left[ \sum_{j:\,\,\, \tau(j) > \tau} 1_{i\sim j} = k\right]
\simeq
\frac{k_0^{1/\alpha}}{k^{1+1/\alpha + o(1)}}
 \ln^{1/\alpha}\left(\frac{t}{\tau}\right).
\end{equation}
In order to obtain the asymptotic probability degree distribution
for an arbitrary node as $t\to \infty$, it is
necessary to sum
(\ref{Krueger}) over all $\tau<t$
that gives
\begin{equation}
\label{Krueger2}
P(k) \simeq   \frac 1t \sum_{0<\tau< t} \frac{k_0^{1/\alpha}}
{k^{1+1/\alpha + o(1)}} \ln^{1/\alpha}\left(\frac{t}{\tau}\right)=
\frac{1}{k^{1+1/\alpha + o(1)}}.
\end{equation}
The emergence of the power law (\ref{Krueger2}) demonstrates
that graphs with a scale-free degree distribution
may appear naturally as the result of a simple edge
formation rule based on choices
where the probability to chose a vertex with
affinity parameter $\omega$ is proportional
to the frequency of appearance of that parameter.
If the affinity parameter $\omega$
is itself power law like distributed one could also
use a direct proportionality
to the value $\omega$ to get still a scale free graph.

\subsection{Trade-offs models of urban sprawl creation}
\label{subsec:Sprawls}
\noindent

The Vermont Forum on Sprawl
defines sprawl as
"dispersed development outside of compact urban and village centers
along highways and in rural countryside"
(the quote appears in \cite{Daltons}).
The term urban sprawl generally has
negative connotations due to the
health and environmental
issues that sprawl creates \cite{Sprawl02}.
Residents of sprawling
neighborhoods tend to emit more pollution per
person and suffer more traffic fatalities.
Urban sprawl is a growing
 problem in many
countries and  in the so called
"smart growth" policies,
 which became
popular in US during the 1990s,
trumping such mainstays as education
 and crime in many polls and
contributing to the election of
 anti-sprawl politicians \cite{Sprawl}.

There are currently many indicators of sprawl but the majority
of them are economic or land-use
related\footnote{For instance, low
 population density is an indicator of sprawl.}
rather than intrinsically spatial or
morphological.

One of the early study
devoted to the
significant reduction of the
number of dead-ends in
the urban network,
caused by the merging of the cart path and road
networks,
suggested that the amount of
 "ringiness" \cite{Hillerhanson}
in the system might serve as a good
morphological indicator of sprawl.
In  \cite{Daltons}, it has
been suggested that the differences
between suburban and urban developments
and even sprawl are clearly discernable
by the proportion and
distribution of cycle lengths
in the dual graph of the network.
Typical suburban developments
usually contain a high ratio of
cul-de-sacs, the dead-end
streets with only one inlet,
together with just a few entrances
accessed from central roads.

However, while observing the
highly interlaced patterns of modern
housing subdivisions sprawled
out over rural lands at the fringe
of many urban areas in
USA \footnote{In accordance to the US
Bureau of Census data on Urbanized Areas,
 Over a 20-year period, the 100 largest Urbanized Areas in US
 sprawled out over an additional 14,545 square miles. That was
  more than 9 million acres of natural habitats, farmland and
   other rural space that were covered over by the asphalt,
   buildings and sub-divisions of suburbia. See
    $http://www.sprawlcity.org/hbis/index.html$ for details. }
     and Canada,
 one can see
     that such subdivisions
may contain no cycles at all
      being an arborescence
offering only a few places
       to enter and
exit the development,
causing traffic
to use high volume collector roads.

Probably, the most fascinatingly
reparative morphological element
of urban sprawl is an
individual access from a private households
to the central path. Designating the
individual parking places as
the client nodes
 and the high volume road as
a hub in the dual graph,
we obtain a
{\it star graph}
as the
 typical syntactic motive pertinent to
urban sprawl.

\begin{figure}[ht]
 \noindent
\begin{center}
\epsfig{file=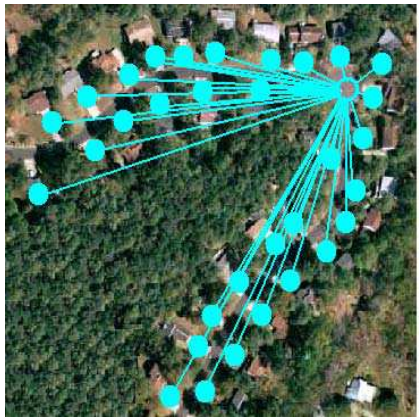,  angle= 0,width =5.5cm, height =5.5cm}
  \end{center}
\caption{\small A segment of Atlanta's suburban sprawl (USA).
The spatial morphology of sprawl is represented by the star graph.}
\label{Fig2_sprawlsstar}
\end{figure}
The dual graph corresponding to the segment of  suburban sprawl shown
in Fig.~\ref{Fig2_sprawlsstar} forms a star graph, in which 32
individual spaces representing the private parking places are
connected to a hub, the only sinuous central road.

Star graphs have been
 observed in many technical
and informational
networks. In particular,
the algorithms generating the tree like
graphs combined
from a number of star subgraphs  have been
extensively studied and
modelled in concern with the Internet
topology (see \cite{InternetGalaxy}-\cite{Fabrikant} and many
others).

Extensive experiments suggest
that  the hierarchical trees containing
the numerous
star subgraphs could arise as a
result of {\it trade-off} process
 minimizing the weighted sums
of two or more objectives.

In the simple model of
Internet growth \cite{Fabrikant},
a tree is built as nodes arrive
 uniformly at random in the unit
square (the shape is, as usual,
 inconsequential). When the $i$-th
 node arrives,
it attaches itself on
one of the previous
 nodes $j$ that minimizes the
weighted sum of the two objectives:
\begin{equation}
\label{costfunction}
\mathrm{cost}(i,j)=\alpha\cdot \mathfrak{d}_{ij}
+\mathfrak{c}_j, \quad \alpha \geq 0,
\end{equation}
where $\mathfrak{d}_{ij}$
 is the  geometrical (Euclidean)
 distance between
 two nodes $i$ and $j$,
and $\mathfrak{c}_j$
is the centrality value
 of  $j$.

It is newsworthy
 that
the behavior of the
trade-off
model
(\ref{costfunction})
 depends
crucially on the
value of the {\it tuning} parameter $\alpha$
which can be naturally
 interpreted as the "last mile" cost reckoning
the construction and maintenance
expenditures.

If $\alpha$ is taken
less than a
particular constant
depending upon the landscape shapes
and
certain economical conditions,
then Euclidean distance
is of no importance, and
 the
network produced by the
trade-off algorithm is
 easily seen to form
a {\it star}.
 A star graph consists of a central node (hub)
characterized by the
 uttermost connectivity
 and a number of terminal
vertices  linked to the hub.

It seems natural to apply the
simple trade-off models
to dual graphs in order
to predict the appearance of urban sprawl
with the local land-use scheme.
The decisive factor for
 the emergence
of star graphs is
the supremacy
of the centrality (integration) objective,
while the physical (Euclidean) distance
between graph vertices
is of no importance.
It
is well known that the
humbleness of physical
distances is among
main factors shaping
 the sprawl land use patterns.
In return,
being highly dependent on
automobiles for transportation,
 the low
density sprawl development consumes
much more land than
traditional urban developments.

\begin{figure}[ht]
 \noindent
\begin{center}
\epsfig{file=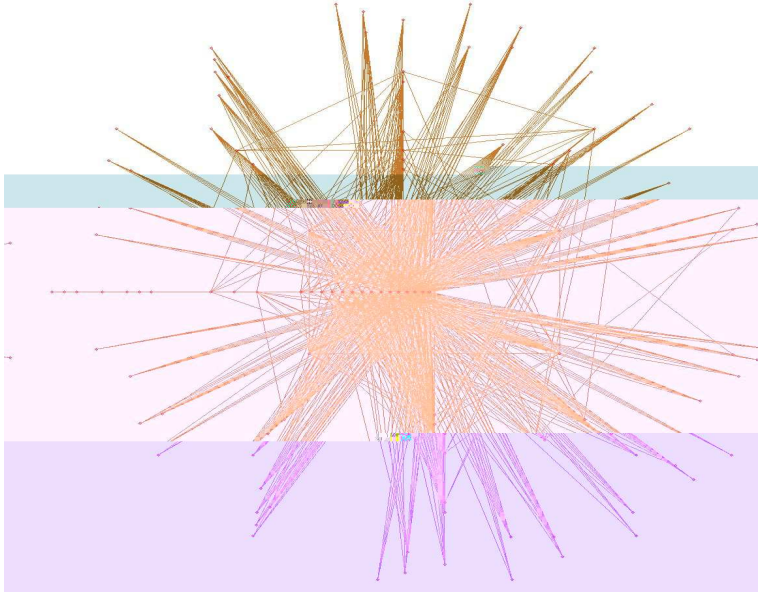,  angle= 0,width =10cm, height =8cm}
  \end{center}
\caption{\small The dual graph of 500 nodes arisen in the course
of trade-off process minimizing two objective functions
simultaneously.} \label{Fig2_03dd}
\end{figure}
The other way around,
if we suggest that
the last mile costs
$\alpha$ grow up
with the network size $N$ at
least as fast as $\sim\sqrt{N}$,
then the
Euclidean distance
between nodes
 becomes an important
factor shaping the form
of the network.
A graph that arises in
the trade-off process
in such a case
constitutes a dynamic
version of the Euclidean
minimum spanning tree, in
which high degrees would
occur, but with exponentially vanishing
probability.

Eventually, if $\alpha$ exceeds a certain
constant, but grows slower than $\sqrt{N}$,
then, almost
certainly, the degrees obey a
power law, and the
resulting dual graph
forms a fractal.

We suggest that the emergence
of a complicated highly inhomogeneous
structure which we observe
in
urban  developments
can generally involve trade-offs, the optimization
 problems between the {\it multiple}, complicated and probably
conflicting objectives.
In order to support this proposition,
 we have
simulated on a
trade-off model
 minimizing {\it two} different objectives simultaneously.
The trading between the geometrical distance and centrality of
nodes has been accounted by the objective (\ref{costfunction}). We
have also used another cost function,
\begin{equation}
\label{cost_2}
\widetilde{cost}(i,j)=\omega\cdot \mathfrak{d}_{ij} + d_{ij}, \quad \omega\geq 0,
\end{equation}
in which $\mathfrak{d}_{ij}$ is the  Euclidean distance and $d_{ij}$
is the graph theoretical distance between nodes $i$ and $j$ (the number
 of hopes), as the second objective.
In general, the structure of graph
that would  arise in such a complex
trade-off process crucially
depends upon the magnitudes of
$\omega$ and $\alpha$, their relative values,
and the way they
depend on the network size $N$.
In general, it may also depend
 upon the local
curvature of the simulation domain
that describes a landscape shape
and its distribution.

In the simplest case of plain domains,
 for the relatively small values of tuning
parameters, $\alpha\simeq \omega
\simeq 0,$
 the geometrical distances between
nodes are of no importance, and  the
structure of the
 resulting graph (shown in \ref{Fig2_03dd})
is shaped by the simultaneous optimization between the graph
distance and  centrality objectives. It is worth mentioning that
the resulting graph sketched in \ref{Fig2_03dd}
 contains a valuable fraction
of twins nodes (while $\alpha= \omega
=10^{-3}$, 84\% of graphs nodes are twins)
forming almost a complete bipartite
 dual graph.

\section{Discussion and Conclusion}
\label{sec:Discussion}
\noindent

Although, nowadays the majority of people live in cities \cite{Crane}
there is no one standard international definition of a city: the term may
be used either for a town possessing city status; for an urban locality
exceeding an arbitrary population size; for a town dominating other towns
with particular regional economic or administrative significance. In most
 parts of the world, cities are generally substantial and nearly always
 have an urban core, but in the US many incorporated areas which have a
  very modest population, or a suburban or even mostly rural character,
   are designated as cities.

A universal social
dynamic underlying the scaling phenomena observed in cities
implies that an increase in productive social opportunities, both
in number and quality,
 leads to quantifiable changes in individual
behavior of humans in a city integrating them into a complex dynamical
network \cite{Macionis}.
The impact of urban landscapes on the construction of
social relations draws attention in
the fields of ethnography, sociology, and anthropology.
In particular, it has been suggested that the urban space combining
social, economic, ideological and technological factors is  responsible
for the technological, socioeconomic, and cultural  development, \cite{Low}.
It is worth to mention that the
processes relating urbanization to economic development and
knowledge production are very general, being shared by all cities
belonging to the same urban system and sustained across different
nations and times.

We suggest that the observed universality 
of the integration and control value statistics 
 we report about in the present paper may help 
to establish the international definition of a city
as a specific land use pattern in which 
streets, places, street junctions and crossroads form an 
 inhomogeneous complex network.
  
\section{Acknowledgment}
\label{Acknowledgment}
\noindent

The work has been supported by the Volkswagen Stiftung (Germany)
in the framework of the project: "Network formation rules, random
set graphs and generalized epidemic processes" (Contract no Az.:
I/82 418). The authors acknowledge the multiple fruitful
discussions with the participants of the workshop {\it Madeira
Math Encounters XXXIII}, August 2007, CCM - CENTRO DE CI\^{E}NCIAS
MATEM\'{A}TICAS, Funchal, Madeira (Portugal).

\newpage
\begin{center}
{\bf \small Table 1: Some features of studied dual city graphs}

\vspace{0.3cm}

\begin{tabular}{c|c|c|c}
   \hline \hline
 Urban pattern   & $N$ & $M$ & $\mathrm{diam}(\mathfrak{G})$ 
    \\ \hline\hline
 Rothenburg ob d.T. & 50 & 115 &  5
 \\
Bielefeld (downtown)& 50 & 142 &  6
 \\ 
 Amsterdam (canals) & 57 & 200 & 7
\\
 Venice (canals) & 96 & 196 &  5
 \\ 
  Manhattan & 355 & 3543 &  5
 \\
  \hline \hline
\end{tabular}
\end{center}


\begin{thebibliography}{000}


\bibitem{Haggett}
P. Haggett, R. Chorley (eds.), {\it Socio-Economic Models
in Geography}, London, Methuen (1967);

 P. Haggett, R. Chorley, {\it Network Analysis in Geography},
  Edward Arnold, London (1969).

\bibitem{Kansky}
 K.J. Kansky, {\it Structure of Transportation Networks: Relationships
Between Network Geometry and Regional Characteristics}, Research Paper {\bf 84},
Department of Geography, University of Chicago , Chicago, IL (1963).


\bibitem{MillerShaw}
H.J. Miller, S.L. Shaw, {\it Geographic Information Systems for
Transportation: Principles and Applications}, Oxford Univ. Press, Oxford (2001).

\bibitem{Batty}
M. Batty, {\it A New Theory of Space Syntax},
UCL Centre For Advanced Spatial Analysis Publications, CASA Working Paper
{\bf 75} (2004).


\bibitem{SSyntax}
 B. Hillier, A. Leaman, P. Stansall,  M.  Bedford,
{\it Environment and Planning B} {\bf 3}, 147-185 (1976).


\bibitem{Hillier96}
 B. Hillier, {\it Space is the machine. A configurational theory of
 architecture},  Cambridge University Press (1996).
\bibitem{Jiang98}
 B. Jiang, "A space syntax approach to spatial cognition in urban environments",
  Position paper for NSF-funded research workshop {\it Cognitive Models of Dynamic
   Phenomena and Their Representations}, October 29 - 31, 1998, University of
   Pittsburgh, Pittsburgh, PA (1998).

\bibitem{Ratti2004}
C. Ratti, {\it Environment and Planning B: Planning and Design},
 vol. {\bf 31}, pp. 487 - 499 (2004).
\bibitem{Krueger89}
 M.J.T. Kruger, {\it On node and axial grid maps: distance
 measures and related topics}. Other. Bartlett School of
 Architecture and Planning, UCL, London, UK (1989).


\bibitem{Hansen59}
 W. G. Hansen, {\it Journal of the
American Institute of Planners} {\bf 25}, 73-76 (1959) .

\bibitem{Wilson70}
A. G. Wilson,  {\it Entropy in Urban and Regional Modelling},
 Pion Press,
London (1970).

\bibitem{glossary}
Bj\"{o}rn Klarqvist, "A Space Syntax Glossary", {\it Nordisk
Arkitekturforskning} {\bf 2} (1993).


\bibitem{JAG}
B. Jiang, Ch. Claramunt, and B. Klarqvist,
{\it JAG}, Vol. {\bf 2} (3/4), 161-171 (2000).


\bibitem{Hillier1987}
 B. Hillier, R. Burdett, J. Peponis,  A. Penn, {\it Architecture
  and Comportement / Architecture and Behaviour} {\bf 3}(3), 233-250 (1987).

\bibitem{ZipfSoo}
Kwok Tong Soo,
 {\it Regional Science and Urban Economics} {\bf 35}(3), 239-263 (2002).

\bibitem{Reed}
W. Reed, {\it Journal of Regional Science}, Vol. {\bf 42}, 1-17 (2002).

\bibitem{Stanley}
H. Maske, A. Halvin., H.E. Stanley, {\it Nature} {\bf 377}, 608-612 (1995).

\bibitem{FractalCities}
M. Batty, P. Longley, {\it Fractal Cities}, Academic Press. London (1994).

\bibitem{Tannier}
C.Tannier, D. Pumain, "Fractals in urban geography : a theoretical outline
and an empirical example",
Cybergeo, article {\bf 307}, 20 (2005)
 ( http://www.cybergeo.presse.fr ).

\bibitem{Carvalho}
R. Carvalho, A. Penn, {\it Physica A} {\bf 332}, 539-547 (2004).

\bibitem{Shinichi}
 B. Hillier, I. Shinichi, {\it Network effects and psychological effects:
  A theory of urban movement}, Proc. of the 5th International Symposium
  on Space Syntax Vol. {\bf 1}, TU Delft,Delft, Netherlands, pp 553–564 (2005).
\bibitem{SSBeijing}
B. Hillier,  "The art of place and the science of space", {\it World Architecture}
{\bf 11}/2005 (185), Beijing, Special Issue on Space Syntax pp. 24-34
 (in Chinese), pp. 96-102 (in English) (2005).



\bibitem{BA}
A.-L. Barab\'{a}si, R. Albert, {\it Science} {\bf 286}, 509-512 (1999).

\bibitem{BA2}
A.-L. Barab\'{a}si, R. Albert, {\it Rev. Mod. Phys.}
{\bf 74}, 47-97 (2002).


\bibitem{CAMEO}
Ph. Blanchard, T. Kr\"{u}ger, The "Cameo Principle" and the Origin of
Scale-Free Graphs in Social Networks, available at arXiv:cond-mat/0302611 (2003).



\bibitem{Simon}
 H. A. Simon, {\it Biometrika} {\bf 42}, 425–440 (1955).

\bibitem{Matthew}
R. K. Merton, {\it Science} {\bf 159}, 56–63 (1968).


\bibitem{Newman2003}
 M.E.J. Newman, {\it SIAM Review} {\bf  45}, 167-256 (2003).
\bibitem{Sprawl02}
J. Norman, H. L. MacLean, and Ch. A. Kennedy,
{\it J. Urban Plannig and Development} {\bf 132}(1), pp. 10-21 (2006).

\bibitem{Sprawl}
{\it A Guide to Smart Growth:
Shattering Myths, Providing Solutions}, Jane S. Shaw and Ronald D.
 Utt, editors; Heritage Foundation, Washington D.C. (2000).

\bibitem{Hillerhanson}
B. Hillier, J. Hanson, {\it The Social Logic of Space} (1993, reprint, paperback
edition ed.). Cambridge: Cambridge University Press (1984).


\bibitem{Daltons}
C. R. Dalton, N.S. Dalton,  "A spatial signature of sprawl: or the
 proportion and distribution of linear network circuits". In:
 {\it GeoComputation 2005}, 1-3 August 2005, Ann Arbor, Michigan, USA (2005).


\bibitem{InternetGalaxy}
M. Castells, {\it The Internet Galaxy: Reflections on the Internet,
 Business, and Society}, Oxford (2001).
\bibitem{Faloutsos}
C. Faloutsos, M. Faloutsos, P. Faloutsos, {\it On power-law relationships
 of the internet topology}, in {\it Proc. CIGCOMM} (1999).
\bibitem{Fabrikant}
A. Fabrikant, E. Koutsoupias, and C.H. Papadimitriou. {\it Heuristically
 optimized tradeoffs: A new paradigm for powerlaws in the internet}. ICALP (2002).

\bibitem{Crane}
P. Crane, A. Kinzig,  {\it Science} {\bf 308}, 1225 (2005).


\bibitem{Macionis}
 J.J. Macionis,  V.N. Parillo,  {\it Cities and Urban Life}, Pearson Education,
Upper Saddle River, NJ (1998).
\bibitem{Low}
S. Low, L. Zuniga (eds.), {\it The Anthropology of Space and Place: Locating
Culture}, Blackwell Publishing (2003).


\end{thebibliography}
\end{document}